\input{style/arxiv-general.cfg}
\documentclass[aoas,MSNbibl,nameyear,seceqn,dvips]{arximspdf}
\makeatletter
   \@ifpackageloaded{graphicx}{}{\usepackage{graphicx}}
\makeatother
\usepackage{multirow}
\usepackage{graphicx}
\usepackage{url,breakurl}

\doi{10.1214/14-AOAS792}
\volume{9}
\issue{1}
\pubyear{2015}
\firstpage{300}
\lastpage{323}
\docsubty{FLA}

\makeatletter
\newcommand{\rrVert}{\Vert}
\newcommand{\llVert}{\Vert}
\newcommand{\eqref}[1]{(\ref{#1})}

\def\bfb{\mathbf{b}}

\def\bfA{\mathbf{A}}
\def\bfC{\mathbf{C}}
\def\bfX{\mathbf{X}}
\def\bfY{\mathbf{Y}}
\def\bfZ{\mathbf{Z}}

\newtheorem{theorem}{Theorem}[section]
\newtheorem{proposition}[theorem]{Proposition}

\newproclaim{definition}[theorem]{Definition}
\newproclaim{assumption}[theorem]{Assumption}

\newproclaim{remark}[theorem]{Remark}

\newcommand{\R}{\mathbb{R}}

\def\toinf{\to\infty}

\makeatother

\begin{document}
\begin{frontmatter}

\title{Inferring gene--gene interactions and functional modules using
sparse canonical correlation analysis\thanksref{T1}}
\runtitle{Inferring gene association networks using SCCA}
\thankstext{T1}{Supported in part by Grants NIH EY019094, NIH U01 HG007031, NSF
DMS-06-36667 and
NSF DMS-11-60319.}

\begin{aug}
\author[A]{\fnms{Y. X. Rachel} \snm{Wang}\ead[label=e1]{rachelwang@stat.berkeley.edu}},
\author[B]{\fnms{Keni} \snm{Jiang}\ead[label=e2]{kenij@berkeley.edu}},
\author[B]{\fnms{Lewis J.} \snm{Feldman}\ead[label=e3]{ljfeldman@berkeley.edu}},
\author[A]{\fnms{Peter~J.}~\snm{Bickel}\ead[label=e4]{bickel@stat.berkeley.edu}}
\and
\author[A]{\fnms{Haiyan} \snm{Huang}\corref{}\ead[label=e5]{hhuang@stat.berkeley.edu}}
\runauthor{Y. X. R. Wang et al.}

\affiliation{University of California, Berkeley }
\address[A]{Y. X. R. Wang\\
P. J. Bickel\\
H. Huang\\
Department of Statistics\\
University of California, Berkeley\\
Berkeley, California 94720\\
USA\\
\printead{e1}\\
\phantom{E-mail:\ }\printead*{e4}\\
\phantom{E-mail:\ }\printead*{e5}}

\address[B]{K. Jiang\\
L. J. Feldman\\
Department of Plant and Microbial Biology\\
University of California, Berkeley\\
Berkeley, California 94720\\
USA\\
\printead{e2}\\
\phantom{E-mail:\ }\printead*{e3}}
\end{aug}

%
\received{\smonth{9} \syear{2014}}

%
\begin{abstract}
Networks pervade many disciplines of science for analyzing complex
systems with interacting components. In particular, this concept is
commonly used to model interactions between genes and identify closely
associated genes forming functional modules. In this paper, we focus on
gene group interactions and infer these interactions using appropriate
partial correlations between genes, that is, the conditional
dependencies between genes after removing the influences of a set of
other functionally related genes. We introduce a new method for
estimating group interactions using sparse canonical correlation
analysis (SCCA) coupled with repeated random partition and subsampling
of the gene expression data set. By considering different subsets of
genes and ways of grouping them, our interaction measure can be viewed
as an aggregated estimate of partial correlations of different orders.
Our approach is unique in evaluating conditional dependencies when the
correct dependent sets are unknown or only partially known. As a
result, a gene network can be constructed using the interaction
measures as edge weights and gene functional groups can be inferred as
tightly connected communities from the network. Comparisons with
several popular approaches using simulated and real data show our
procedure improves both the statistical significance and biological
interpretability of the results. In addition to achieving considerably
lower false positive rates, our procedure shows better performance in
detecting important biological pathways.
\end{abstract}

%
\end{frontmatter}

\section{Introduction}
\label{sec_intro}
Many complex systems in science and nature are composed of interacting
parts. Such parts can be modeled as nodes and their relationships as
edges in a network. Network modeling has found numerous applications
[\citet{Newman:2010}]. Gene association networks is one such example,
with genes modeled as nodes and their interactions as edges. One
important application of gene networks is the identification of
communities corresponding to genes with related functional groupings.
Many of these functional groups encode biological pathways. A major
task in understanding biological processes is to identify these pathway
genes and elucidate the relationships between them. We focus in this
paper on modeling gene interactions. As a result, a gene network can be
constructed using the interaction measures as edge weights and gene
functional groups can be inferred as tightly connected communities in
the network.

In gene networks, direct observation of gene relationships by
experimental approaches is extremely cost-prohibitive given that the
typical size of the networks is in the tens of thousands. The gene
expression levels, on the other hand, are easier to measure and can be
regarded as sets of covariates associated with the nodes. Constructing
gene networks using expression data has remained a challenging
unsupervised learning problem in the statistics literature due to the
complexity of data structure and the difficulty of finding an
appropriate measure for characterizing gene relationships. A review of
existing methods can be found in \citet{Wang:2014}.

Most methods for inferring edges in gene networks are based on the
notion of measuring expression profile similarity or co-expression,
which aims to estimate marginal relationships between pairs of genes.
Widely used co-expression measures include the Euclidean distance or
the angle between vectors of observed expression levels or, most
commonly, the marginal covariance or correlation. Measures detecting
general statistical dependence such as mutual information (MI) are also
explored. MI offers the advantage of being able to detect nonlinear
correlations [\citet{Daub:2004}], but some empirical studies [\citet
{Steuer:2002}] also show it yields almost identical results as the
Pearson correlation. Recently, a new measure named the maximal
information coefficient (MIC) was proposed by \citet{Reshef:2011} based
on normalized estimates of MI. \citet{Kinney:2014} offer some criticisms
and discussions of MIC.

The above measures for estimating marginal dependencies only consider
pairwise relationships. However, in a real biological pathway, a gene
can interact with a group of genes but their marginal relationships may
remain weak. Such higher-level interactions (i.e., gene group
interactions) are better modeled by Gaussian graphical models (GGM) due
to its interpretation in terms of conditional correlations. Under the
assumption of multivariate normality of gene expression vectors, the
GGM uses the inverse of the gene covariance matrix (or precision
matrix) as a measure for gene associations. This approach is closely
related to the concept of partial correlations: the $(i,j)$th element
in the precision matrix is proportional to the partial correlation
between gene $i$ and $j$ conditional on the rest of the genes. To
address the ``curse of dimensionality'' (the number of genes being much
larger than the number of samples) in estimating the precision matrix,
one can exploit the belief that gene networks are inherently sparse and
reframe the problem of estimating partial correlations in a penalized
regression setting [\citet{Meinshausen:2006,Peng:2009}]. More studies
on estimating the sparse precision matrix in high-dimensional GGMs can
be found in, for example, \citet{Schafer:2005a}, \citet{Friedman:2007}
and \citet{Zhou:2011}.

Despite their attractive theoretical properties, these partial
correlation-based methods still have limitations in their estimation
methods. In the current literature, partial correlation is usually
calculated conditioned on either all of the available genes or a more
or less arbitrary subset of them that may contain noisy (biologically
unrelated) genes. \citet{Fuente:2004} reported that conditioning on all
genes simultaneously can introduce spurious dependencies which are not
from a direct causal or common ancestors effect. To alleviate this
concern, there are alternative approaches using lower order partial
correlations [\citet
{Li:2002,Fuente:2004,Magwene:2004,Wille:2004,Wille:2006}] which
condition on one or two other genes. However, these
methods come at a cost of lowering the sensitivity for inferring higher
level gene associations and do not necessarily eliminate the effect of
noisy genes. \citet{Kim:2012} proposed to minimize the impact of noisy
genes by conditioning on a small set (3--5 genes) of ``seed genes''
(i.e., known pathway genes). However, such prior biological information
is not always available, especially in exploratory studies.

In this paper we tackle the problem of estimating gene relationships
when the correct conditional set for partial correlation is unknown. We
introduce a new method of inferring the strength of gene group
interactions using sparse canonical correlation analysis (SCCA) with
repeated random partition and subsampling of the gene expression data
set. There has been a growing interest in applying SCCA to genomic data
sets [\citet{Waaijenborg:2008,Parkhomenko:2009,Witten:2009b,Lee:2011}]
in the context of studying relationships between two or more
sets of variables, such as gene expression levels, copy numbers and
other phenotype variations, with measurements taken from the same
sample. One novelty of our method lies in the application of SCCA to a
single data set facilitated by a random partition scheme. By randomly
separating the genes into two groups, SCCA searches for a strong linear
relationship between a small set of genes, for example, 5--20 genes,
from both groups of genes (e.g., 500--2000 genes in total). Through
multiple rounds of random partition, this SCCA approach, reframed in a
linear regression setting, gives estimates proportional to partial
correlations conditioned on different sets of signal genes (with noisy
genes eliminated through sparsity). The subsampling procedure analyzes
different subsets of the genes at a time and enables simultaneous
identification of multiple interacting groups with different signal
strengths. Using this construction, we build an edge weight matrix for
the whole gene network whose interaction measure reflects an aggregated
estimate of partial correlations of different orders. Our approach is
flexible and can be adapted to work with or without prior biological knowledge.

The rest of the paper is organized as follows. In Section~\ref
{sec_methods} we discuss in detail the motivations behind our new
scheme of computing edge weights in a gene network by assessing gene
group interactions and provide an outline of the full procedure. To
identify densely connected communities as potential gene functional
modules in the constructed network, we implemented two well-known
methods in the network literature, the stochastic block model (SBM) and
hierarchical clustering (HC). In Section~\ref{sec_results} comparisons
are made between our procedure and correlation-based methods.
We demonstrate that our procedure in general achieves a significant
reduction in the rate of false positives. To test its performance in
real data applications, our procedure is applied to an \textit
{Arabidopsis thaliana} microarray data set obtained under oxidation
stress. Finally, in Section~\ref{sec_discussion} we discuss the
advantages and potential extensions of the present method.

\section{Methods}
\label{sec_methods}
As mentioned in Section~\ref{sec_intro}, the conditional correlation
interpretation of partial correlation suggests it is a more appropriate
framework for modeling higher level interactions in gene networks,
provided the conditional computation is carried out properly. In this
section, we discuss some of the limitations of the partial correlation
approach that arise due to its reliance on the correct selection of
conditional sets of genes and how our SCCA-based approach circumvents
this difficulty. We then give a detailed description of our new method
of estimating an edge weight matrix using SCCA with subsampling. 

\subsection{Method motivation}
Recall that when the gene expression levels follow a multivariate
normal distribution, for a set of genes $W$, the partial correlation
between genes $i$ and $j$ can be expressed as
%
\begin{equation}
\rho_{ij} = \operatorname{cor}\bigl(i,j|W\setminus\{i,j\}\bigr)=
\cases{ \displaystyle -\frac{\omega_{ij}}{\sqrt{\omega_{ii}\omega_{jj}}},
&\quad $i\neq j,$ \vspace*{2pt}
\cr
1, &\quad $i=j$,}
\label{eq_partial_cor}
\end{equation}
where $\omega_{ij}$ are elements in the precision matrix $(\Sigma
^G)^{-1}$ with $\Sigma^G$ being the gene covariance matrix of the set
$W$ [see, e.g., \citet{Edwards:2000}]. Genes $i$ and $j$ being
conditionally independent is equivalent to the corresponding partial
correlation and element in the precision matrix being zero.

As pointed out in \citet{Fuente:2004} and \citet{Kim:2012}, the selection
of a proper set of genes on which the correlation in~\eqref
{eq_partial_cor} is conditioned determines the effectiveness of using
partial correlation to measure gene interactions. The inclusion of
noisy (biologically unrelated) genes in the set $W\setminus\{i,j\}$
may introduce spurious dependencies and, consequently, false edges in
the estimated network. The use of partial correlation may also prove
problematic when $W$ contains multiple pathways. Here is a minimal
example: suppose the set $W$ has two pathways $\{x,y,z\}$ and $\{u,v\}$
and two independent noisy genes $p$ and $q$, with expression relationships
%
\begin{equation}
z = x+y+\varepsilon_1u+\varepsilon_2v+
\varepsilon_3p,\qquad u = \delta _1x+\delta _2y+
\delta_3z+\delta_4q+v, \label{eqn_minExample}
\end{equation}
where $\varepsilon_i$ and $\delta_j$ are small constants so that the
dependencies between the two pathways are negligible, and gene $v$ is
independent of genes $x$ and $y$. Computing the partial correlations,
we have the desired dependencies:
\begin{eqnarray*}
\operatorname{cor}\bigl(z,x|W\setminus\{z,x\}\bigr) & =& \operatorname {cor}
\bigl(z,y|W\setminus\{z,y\} \bigr)=1,
\\
\operatorname{cor}\bigl(u,v|W\setminus\{u,v\}\bigr) & =& 1,
\end{eqnarray*}
but also some spurious ones:
\[
\operatorname{cor}\bigl(u,x|W\setminus\{u,x\}\bigr)=\operatorname {cor}
\bigl(u,y|W\setminus\{u,y\} \bigr)=\operatorname{cor}\bigl(u,z|W\setminus\{u,z\}
\bigr)=1.
\]
Using these partial correlations to construct an edge weight matrix
would imply the two pathways are fully connected. The proper
calculation should condition only on genes in the same pathway, but
such information is usually hard to obtain in practice. Alternatively,
a more appropriate edge weight measure can take into account the
magnitude of the linear coefficients in~\eqref{eqn_minExample} so that
it reflects the amount of contribution each gene makes to a pathway and
the two-block nature of the network. Recall that in a regression
setting, the regression coefficients are multiplicative functions of
the corresponding partial correlations. In this sense, the coefficients
encompass more information and provide a better resolution on gene
relationships than the partial correlations alone.

Motivated by these observations, we propose a new way to assess gene
group interactions. In particular, we aim to identify strong linear
relationships possessed by a small subset of the candidate genes. We
make direct use of the linear coefficients found by SCCA when applied
to two randomly partitioned gene groups. With repeated random partition
on subsampled gene sets, an edge weight matrix built by the average
SCCA coefficients over iterations reflects an aggregated level of
direct or partial gene interactions. More discussion on how CCA
coefficients relate to partial correlations can be found in Section~4
of the supplementary information [\citet{supp}]. Sparsity is imposed to reduce
dimensionality and, in particular in the example above, ensures the
mixing of the two pathways is negligible on average. 

\subsection{Review of sparse canonical correlation analysis and its
implementation}
Let $\bfX\in\R^{n\times q_1}$ be a matrix comprised of $n$ observations
on $q_1$ variables, and $\bfY\in
\R^{n\times q_2}$ a matrix comprised of $n$ observations on $q_2$
variables. CCA introduced by \citet{Hotelling:1936} involves finding
maximally correlated linear combinations between the two sets of
variables. More explicitly, one finds
$\bolds{\alpha}\in\R^{q_2}$ and $\bolds{\beta}\in\R^{q_1}$
that solve
the optimization problem
%
\begin{equation}
\max_{\bolds{\alpha},\bolds{\beta}} \bolds{\alpha}^T\Sigma _{YX}
\bolds{\beta} \qquad\mbox{subject to } \bolds{\alpha}^T\Sigma
_{YY}\bolds{\alpha}=1,\bolds{\beta}^T\Sigma_{XX}
\bolds{\beta}=1, \label{eq_CCApop}
\end{equation}
where $\Sigma_{(\cdot,\cdot)}$ represent the correlation matrices. Note
that provided the variables in $\bfX$ and $\bfY$ have nonzero
variances, this is equivalent to the usual CCA formulation in terms of
covariance matrices.

In practice, the population correlations are replaced with their sample
counterparts. That is, $S_{YX}=\bfY^T\bfX/(n-1)$, $S_{XX}=\bfX^T\bfX
/(n-1)$ and $S_{YY}=\bfY^T\bfY/(n-1)$, assuming the columns of $\bfX$
and $\bfY$ have been centered and scaled. Let ${\mathbf{a}}$ and
${\bfb}$ be
the weight vectors solving the optimization problem
%
\begin{equation}
\max_{\mathbf{a},\bfb}\mathbf{a}^T S_{YX} \bfb\qquad
\mbox{subject to }\mathbf{a}^T S_{YY} \mathbf{a}=1,
\bfb^T S_{XX} \bfb=1 \label{eq_CCAsample}
\end{equation}
for sample correlations.

For high throughput biological data, $q_1$ and $q_2$ are typically much
larger than $n$. It is thus natural to impose sparsity on $\mathbf{a}$ and
$\bfb$, and this can be done by including (typically convex) penalty
functions in~\eqref{eq_CCAsample}. A number of studies [\citet
{Waaijenborg:2008,Witten:2009a,Parkhomenko:2009}] have proposed
various methods for formulating the penalized optimization problem and
obtaining sparse solutions. Here we adopt the diagonal penalized CCA
criterion given by \citet{Witten:2009a}, which treats the covariance
matrices in~\eqref{eq_CCAsample} as diagonal and relaxes the equality
constraints for convexity:
%
\begin{equation}\qquad
\max_{\mathbf{a},\bfb}\mathbf{a}^T\bfY^T \bfX\bfb\qquad
\mbox{subject to }\mathbf{a}^T\mathbf{a} \leq1, \bfb^T\bfb
\leq1, p_1(\mathbf{a})\leq c_1, p_2(\bfb)
\leq c_2, \label{eq_SCCA}
\end{equation}
where $p_1$ and $p_2$ are convex penalty functions. In this paper, we
consider an $L_1$ penalty and solve the above optimization using the
modified NIPALS algorithm proposed by \citet{Lee:2011}, which is
reported to yield better empirical performance than
Witten, Tibshirani and
Hastie's
(\citeyear{Witten:2009a}) algorithm. The modified NIPALS algorithm performs penalized
regressions iteratively on $\bfX$ and $\bfY$ with the penalty functions
$p_{\lambda_1}(\cdot)=\lambda_1\llVert \cdot\rrVert _1$ and
$p_{\lambda_2}(\cdot)=\lambda_2\llVert \cdot\rrVert _1$.
This is
an equivalent formulation to iteratively optimizing~\eqref{eq_SCCA}
using the bounded constraints.

It is important to note that one more complication arises when SCCA is
applied to gene expression data. In CCA, the estimation of the
correlation matrix using sample correlations requires the data matrices
$\bfX$ and $\bfY$ have independent rows. However, given a gene
expression matrix with genes in columns and experiments in rows, it is
often the case that row-wise and column-wise dependencies co-exist.
Row-wise dependencies, or experiment dependencies, can be defined as
the dependencies in gene expression between experiments due to the
similar or related cellular states induced by the experiments [\citet
{TengHuang:2009}]. When unaccounted for, they can introduce
redundancies that overwhelm the important signals and lead to
inaccurate estimates of the gene correlation matrix. To decouple the
effect of experiment dependencies from the estimation of gene
correlations, we apply the \textit{Knorm} procedure from \citet
{TengHuang:2009}. The \textit{Knorm} model assumes a multiplicative
structure for the gene--experiment interactions, and iteratively
estimates the gene covariance matrix and experiment covariance matrix
through a weighted correlation formula. In addition, row subsampling
and covariance shrinkage are used to ensure robust estimation.

\subsection{Constructing an edge weight matrix by SCCA with repeated
random partition and subsampling}
\label{subsec_SCCAsub}
Suppose an observed data set contains measurements of the expression
levels of $p$ genes in $n$ experiments, where each experiment has a
small number of replicates. We next describe our new procedure of
computing edge weights that reflect gene group interactions in the gene
network.\vadjust{\goodbreak}

\textit{Summary of procedure}:

\textit{Step} (i): \textit{Data normalization by Knorm}. A gene expression
matrix $\bfZ_b$ of dimension $n\times p$ can be generated from the full
data set by sampling one replicate from each experiment. Using the
\textit{Knorm} model in \citet{TengHuang:2009}, we normalize $\bfZ_b$ as
%
\begin{equation}
\bfZ^*_b=\bigl(\hat{\Sigma}^E\bigr)^{-1/2}(
\bfZ_b-\hat{\mathbf M}), \label{eq_normalized_by_Knorm}
\end{equation}
where $\hat{\mathbf M}$ is the estimated mean matrix and $\hat{\Sigma}^E$
is the estimated experiment correlation matrix.

\textit{Step} (ii): \textit{Subsampling.} For each normalized expression
matrix $\bfZ^*_b$, sample (without replacement) a fixed fraction $s$,
say, 70\%, of the genes to obtain an $n\times sp$ submatrix $\bfZ
^{\operatorname
{sub}}_{b}$.

\textit{Step} (iii): \textit{SCCA with random partition on the subsampled
matrix}. For each partition $t$, randomly split the columns (genes) of
$\bfZ^{\operatorname{sub}}_{b}$ into two groups of equal size (more
explanation
given in the remarks below) to form $\bfX^{\operatorname{sub}}_{b,t}$
and $\bfY
^{\operatorname{sub}}_{b,t}$. Run SCCA on $\bfX^{\operatorname
{sub}}_{b,t}$ and $\bfY
^{\operatorname{sub}}_{b,t}$: find sparse weight vectors $\mathbf
{a}^{\operatorname
{sub}}_{b,t}$ and $\bfb^{\operatorname{sub}}_{b,t}$ using the
modified NIPALS
algorithm [\citet{Lee:2011}] with the $L_1$ penalty and tuning
parameters $\bolds{\lambda}=(\lambda_1,\lambda_2)$, the choice of
which will be discussed in Section~\ref{sec_results}.

Let $\mathbf{c}_{b,t}$ be the list of the absolute values $|\mathbf
{a}^{\operatorname
{sub}}_{b,t}|$ and $|\bfb^{\operatorname{sub}}_{b,t}|$ ordered
according to the
gene list. For the genes not included in the subsampled matrix, the
corresponding values in $\mathbf{c}_{b,t}$ are set to $0$. Average
over all
the partitions to obtain the average weights $\bar{\mathbf{c}}_b$.
Define edge
weight matrix $\bfA_b=\bar{\mathbf{c}}_b\bar{\mathbf{c}}_b^T$,
setting $\operatorname
{diag}({\bfA}_b)=0$ to exclude self loops. 


\textit{Step} (iv): \textit{Repeat steps \textup{(ii)} and \textup{(iii)} $B$ times}. Define
$\bar{\bfA}=1/B\sum_{b=1}^B {\bfA}_b$ and normalize by the maximum
value in $\bar{\bfA}$.

As will be demonstrated in Section~\ref{subsec_simulation}, $\bar
{\bfA}$
defined above exhibits a natural block structure when there is one or
multiple functional groups. Here are more remarks on our procedure to
construct $\bar{\bfA}$:
\begin{longlist}[1.]
\item[1.] Step (i) can be skipped when dependencies between experimental
conditions are weak and not of concern.

\item[2.] Step (ii) subsampling is necessary if we aim to identify multiple
functional groups (that may overlap) simultaneously. As there will be
multiple groups with strong interactions, not all of them can be
detected unless different subsets of genes are considered. For more
discussion about the subsampling step and the choice of subsampling
levels, we refer to Section~5 in the supplementary information [\citet{supp}].

\item[3.] During the random partition in step (iii), the two sets of genes
do not have to be exactly equal in size, but they need to be comparable
in order to maximize the chance of separating any gene functional group
of interest into two sets.

\item[4.] Through multiple rounds of random partition, SCCA gives estimates
in a regression setting proportional to partial correlations
conditioned on different sets of signal genes. Overall, subsampling and
random partition enable us to consider different subsets of the genes
and ways to group them. Thus, the elements in $\bar{\bfA}$ can be
interpreted as an aggregated measure of partial correlations of
different orders as the algorithm steps through different conditional
sets of genes.

\item[5.] As we search through different subsets of genes, different signal
groups are identified depending on the strengths of linear associations
in the subset. As will be shown empirically in Section~\ref
{subsec_simulation}, the averaged result leads to the formation of a
distinct block structure with different connectivities in the matrix.
\end{longlist}

Our procedure is flexible and can be modified easily to incorporate the
following variants:
\begin{longlist}[1.]
\item[1.] If prior knowledge is available on a pathway of interest, for
example, it is known in advance that some genes are actively involved
in that pathway, one may focus on the identification of the gene group
related to this pathway first and incorporate the prior knowledge by
lowering the penalties associated with those known pathway genes in the
SCCA algorithm. Examples involving using prior knowledge of pathway
genes can be found in Section~5 of the supplementary information [\citet{supp}].

\item[2.] If the interest is to identify disjoint gene groups and running
time is not a concern, we can run the whole procedure iteratively with
no subsampling, each time identifying one dominating signal group and
removing it from the subsequent analysis.
\end{longlist}

\textit{Asymptotic behavior of our procedure.}
Here we first show asymptotically the validity of our procedure by
considering a simple case where there exists only one functional group
and all the other genes are uncorrelated. Due to this simplification,
no subsampling is needed, and the use of CCA without sparsity suffices
since in the asymptotics we consider the regime of $n$ (number of
experiments) going to infinity with $p$ (number of genes) fixed.
Without loss of generality, in the entire gene set $G=\{1,2,\ldots,p\}$
let the first $k$ genes $K=\{1,2,\ldots,k\}$ form one pathway.

For every partition $t$, let $\mathbf{a}_t$ and $\bfb_t$ be the solutions
to~\eqref{eq_CCAsample} and ${\mathbf{c}}_t$ be the list of the absolute
values $|\mathbf{a}_t|$ and $|\bfb_t|$ ordered according to the gene list.
Assuming $\bfZ$ follows a multivariate normal distribution and the
inverse covariance matrix has a diagonal block structure (detailed
assumptions are presented in Section~2 of the supplementary
information [\citet{supp}]), we have the following proposition regarding the
asymptotic difference between the values of $\{c_{i,t},i\in K\}$ and $\{
c_{j,t},j\notin K\}$ averaged over $t$. For convenience suppose $p$ is
even and denote $q=p/2$.

\begin{proposition}
\label{prop_asympDiff}
Let $\bar{\mathbf{c}}=\sum_{t=1}^{N}\mathbf{c}_t/N$, where $N$ is
the number of
partitions, then given $1<k<q$,
%
\begin{equation}
\lim_{N\toinf}\lim_{n\toinf} \Bigl(\min
_{i\in K}\bar{c}_i-\max_{j\notin
K}
\bar{c}_j\Bigr) = D
\end{equation}
for some positive constant $D$.
\end{proposition}
In Section~2 of the supplementary information [\citet{supp}], we give the proof of
Proposition~\ref{prop_asympDiff} with a lower bound on $D$ that
quantifies the asymptotic difference in the assigned weights between
functional group genes and noisy genes. The separation in $\bar
{\mathbf{c}}$
implies the genes in the graph characterized by the edge weight matrix
$\bar{\bfA}=\bar{\mathbf{c}}\bar{\mathbf{c}}^T$ can be grouped
into different
clusters based on their connectivity.

To further understand the asymptotic behavior of our procedure in
general cases when multiple functional groups exist, we present an
example that consists of two (disjoint) groups of interacting genes and
other unrelated genes in supplementary information Section~2 [\citet{supp}]. We show a
theoretical derivation of $\bar{\bfA}=1/B\sum_{b=1}^B {\bfA}_b =
1/B\sum_{b=1}^B \bar{\mathbf{c}}_b\bar{\mathbf{c}}_b^T$ for this example
in detail to
highlight and explain the role of subsampling. We can see that with
subsampling, the limiting $\bar{\bfA}$ (when $n\to\infty$) exhibits a
natural block structure corresponding to the two gene groups, thus
extending the validity of Proposition~\ref{prop_asympDiff}. The ideas
underlying the analytical derivation in this simple example are
straightforward and directly applicable to general cases, though the
computations involved would be very tedious. Note that the analytical
computations look tedious even in this small example.


\subsection{Identify community structures given the edge weight matrix
\texorpdfstring{$\bar{\mathbf{A}}$}{barA}}
\label{subsec_clustering}

To demonstrate that $\bar{\bfA}$ possesses advantages over traditional
approaches in identifying gene functional modules, subsequent analysis
of $\bar{\bfA}$ based on community detection tools is needed. Many
methods are available in this field. In particular, clustering has been
a popular and well-studied technique. \citet
{Kaufman:2005,Theodoridis:2008,Jain:1999} provide general reviews of various
clustering techniques, and reviews with more specific focus on gene
expression data can be found in \citet
{D'haeseleer:2000,Jiang:2004,Kerr:2008}. Variants of spectral
clustering are also widely explored
for detecting communities in sparse networks [\citet{Ramesh:2010}].
Viewing gene relationships as edges in a graph, a natural approach is
to consider functional modules as tightly connected subgraphs. Genes
with related functionalities are expected to have dense connections,
whereas biologically unrelated (noisy) genes may be only sparsely
connected. The Stochastic Block Model (SBM) builds a general
probabilistic graph model based on such an assumption that nodes
(genes) have different connectivities depending on their block memberships.

Below we introduce two popular community detection tools, SBM and
hierarchical clustering (HC), which we will use in later simulation and
real data analysis to dissect gene interaction groups from $\bar{\bfA}$.
As we have mentioned, there are many other choices for performing this
task. The structure of $\bar{\bfA}$ itself may also imply some methods
are more suitable than others. In this paper, it is not our intention
to suggest or evaluate the best community detection tools that should
be applied to~$\bar{\bfA}$. Here we are presenting SBM and HC just as two
illustrative approaches.

The SBM, formally introduced by \citet{Holland:1983}, generalizes the
Erd\H{o}s--R\'{e}nyi model and defines a family of probability
distributions for a graph. Here is a detailed model definition.

\begin{definition}
A SBM is a family of probability distributions for a graph with node
set $\{1,2,\ldots,p\}$ and $Q$ node blocks defined as follows:
\begin{longlist}[1.]
\item[1.]
Let $\bfC=(C_1,C_2,\ldots,C_p)$ denote the set of labels such that
$C_i=k$ if the node $i$ belongs to block $k$:
\[
\bfC\stackrel{\mathrm{i.i.d.}} {\sim} \operatorname{Multinomial} ({\bolds \gamma}),
\]
where ${\bolds\gamma}=(\gamma_1,\gamma_2,\ldots,\gamma_Q)$ is the
vector of porportions.
\item[2.]
Let $\bolds\pi= (\pi_{lk})_{1\leq l , k \leq Q}$ be a symmetric
matrix of a block dependent edge probability matrix and $\bfA$ be the
adjacency matrix. Conditioned on the block labels $\bfC$, $(\bfA_{ij})$
for $i<j$ are independent, and
\[
P(\bfA_{ij}|\bfC)=P(\bfA_{ij}=1|C_i=l,C_j=k)=
\pi_{lk}.
\]
\end{longlist}
\end{definition}
Discretizing $\bar{\bfA}$ defined in Section~\ref{subsec_SCCAsub}
into a
0--1 matrix, the class labels and the parameters $\bolds\gamma$ and
$\bolds\pi$ are estimated using the psuedo-likelihood algorithm by
\citet{Amini:2013}. The unconditional version of the algorithm fits the
conventional SBM above, while the conditional version takes into
account the variability of node degrees within blocks [\citet
{KarrerNewman:2010}]. Potential functional groups are identified as
classes having large diagonal entries in $\bolds\pi$.

Agglomerative HC is another widely used nonmodel-based technique for
extracting communities, especially in the study of social networks
[\citet{Scott:2000}]. In our application, we adopt Ward's distance
[\citet
{Ward:1963}] for the computation of merging costs. Let $g_i$ be the
nodes, the distance between two clusters $M_1$, $M_2$ defined as
\begin{eqnarray*}
d(M_1,M_2) & = &\frac{n_1 n_2}{n_1+n_2}\llVert
m_1-m_2\rrVert ^2
\\
& =& \frac{1}{2(n_1+n_2)}\sum_{i,j\in M_1\cup M_2}\llVert
g_i-g_j\rrVert ^2 - \frac{1}{2n_1}\sum
_{i,j\in M_1}\llVert g_i-g_j\rrVert
^2
\\
& &{}-\frac{1}{2n_2}\sum_{i,j\in M_2}\llVert
g_i-g_j\rrVert ^2,
\end{eqnarray*}
where $n_1$ and $n_2$ denote the sizes of $M_1$ and $M_2$, and $m_1$
and $m_2$ are the cluster centers of $M_1$ and $M_2$, respectively. A
natural way to define the square of the pairwise distance is $\llVert  g_i-g_j\rrVert ^2 = 1-\bar{\bfA}_{ij}$ for $i\neq j$, and zero
otherwise. Since Ward's method minimizes the increase in the within
group sum of squares at each merging and tends to merge clusters that
are close to each other and small in size, a~small cluster that manages
to survive a long distance before coalescing is likely to be a tight
cluster, indicating the genes it contains have high connectivity with
each other. Thus, at an appropriately chosen cutoff level $Q$, we
identify the smallest few clusters as potential functional groups.

Both SBM and HC require a priori knowledge of the number of clusters
$Q$, and the proper selection of $Q$ remains an open problem in the
literature. For SBM, we refer to some discussions in \citet{Daudin:2008}
and \citet{Daudin:2012}. For HC, a common way to choose the cutoff $Q$
is to set it as the number just before the merging cost starts to rise
sharply. Due to the scale and complexity of a typical gene expression
data set, this criterion is not very applicable. In this paper, for the
HC approach we choose $Q$ empirically based on the sizes of the
clusters each $Q$ produces. That is, $Q$ is increased incrementally
until small clusters start to emerge. A comparison between SBM and HC
can be found in Section~\ref{subsec_simulation}.

\subsection{Flow chart summarizing the whole procedure}
\begin{figure}

\includegraphics{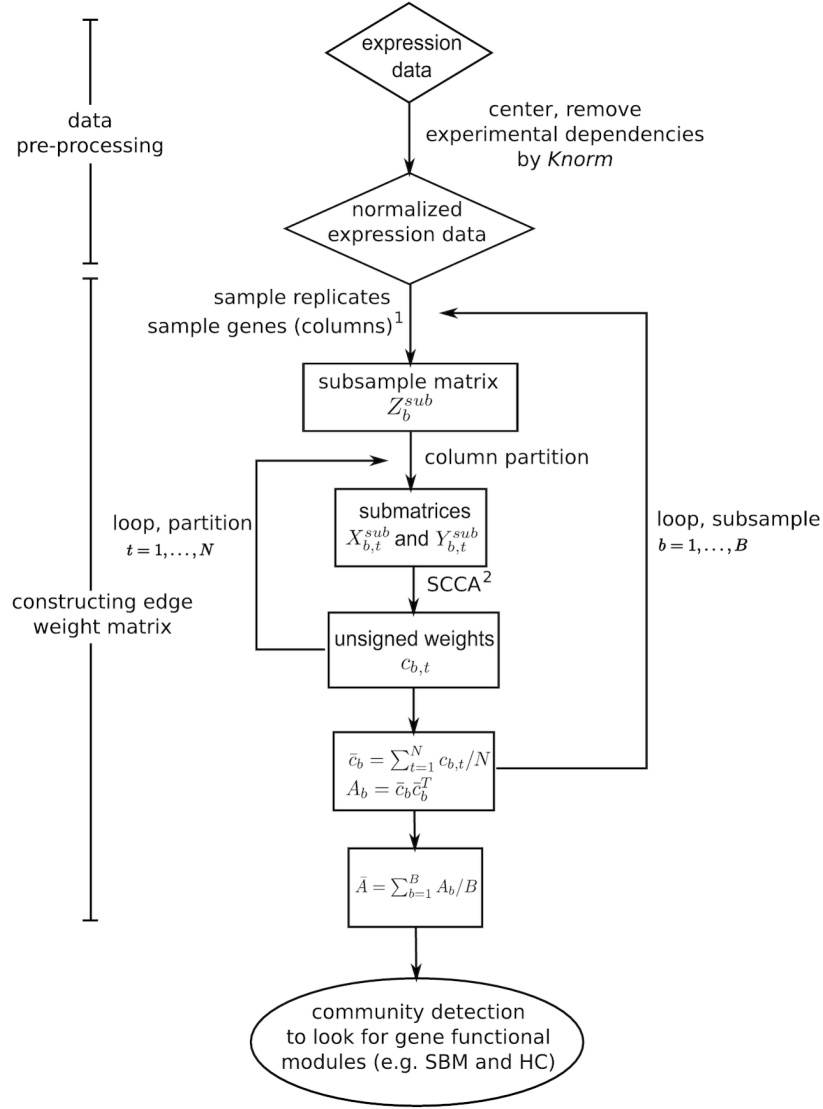}

\caption{Flow chart summarizing the whole procedure. Each numeric
superscript in the diagram indicates the need for tuning parameters: 1.
Subsampling level, 2. Penalty parameter $\bolds\lambda$.}
\label{fig_flowChart}
\end{figure}

A comprehensive summary of the whole procedure, including the tuning
parameters needed in constructing $\bar{\bfA}$ and illustrative
subsequent analysis of $\bar{\bfA}$, is provided in Figure~\ref
{fig_flowChart}. The choices of the parameters are explained in the
paper and summarized again in Section~3 of the supplementary information [\citet{supp}].

\section{Results}
\label{sec_results}
In this section we evaluate the performance of the proposed method and
other approaches using simulated and real microarray data sets. In
particular, we compare the quality of the estimated gene functional
groups, resulting from different ways of computing edge weights, and
the two methods of community detection (SBM and HC) discussed in
Section~\ref{subsec_clustering}. We use \textit{precision} and
\textit
{recall}, defined as $\mathit{precision}=\mathit{TP/(TP+FP)}$ and
$\mathit{recall}=\mathit{TP/(TP+FN)}$, as measures for evaluating
classification performance. Here \textit{TP} is the number of true
positive findings of functional group genes, \textit{FP} is the number
of false positives and \textit{FN} is the number of false negatives. In
the context of this study, they can be regarded as a measure of
exactness and completeness of our search results, respectively. The
problems of choosing the appropriate proportion of subsampling and
$\bolds\lambda$ for sparsity are also discussed. For detailed analysis
of the effects of the tuning parameters, we refer to Section~\ref
{sec_sensitivity} and Section~7 of the supplementary information [\citet{supp}].

\subsection{Simulation}
\label{subsec_simulation}

\subsubsection{Generation of simulation data sets}
We simulate a microarray data set consisting of $p=150, 300 \mbox{ or }
500$ genes and $n=30$ experiments, with 5 replicates for each
experiment. To make the data more realistic, we introduce experiment
dependencies,
multiple functional groups and random noise. The simulation parameters
are generated as follows:
\begin{longlist}[(ii)]
\item[(i)] Experiment correlation matrix, $\Sigma^E$. For illustrative
purpose, we set the experiment correlation matrix to have 0, 33 and
67\% dependencies. In the case of a 33\% dependency, for example, 33\%
of the experiments have high dependencies (correlation between 0.5 and
0.6) while the remaining experiments are uncorrelated with one another.

\item[(ii)] Gene correlation matrix, $\Sigma^G$. In each data set, we
introduce one or two functional groups with 15 genes in each. Genes in
the same group are correlated, having either high correlations (0.5--0.6) or low correlations (0.1--0.2) with the other genes, and
otherwise they are not.
\end{longlist}

Using the above parameters, we generate the simulation data as follows.
First, we generate
a $30\times500$ gene expression matrix $\bfZ$, with $\operatorname{vec}(\bfZ^T)$,
from a
multivariate normal distribution with mean zero and
a covariance matrix $\Sigma^G\otimes\Sigma^E$. To introduce linear
relationships, within each group we take linear combinations of some
genes to replace their original values. Using the final $30\times500$
gene expression matrix, we add random
noise with a small SD (e.g., 0.01) to each row to
generate the 5 replicates for each experiment.

\begin{figure}
\centering
\begin{tabular}{@{}cc@{}}

\includegraphics{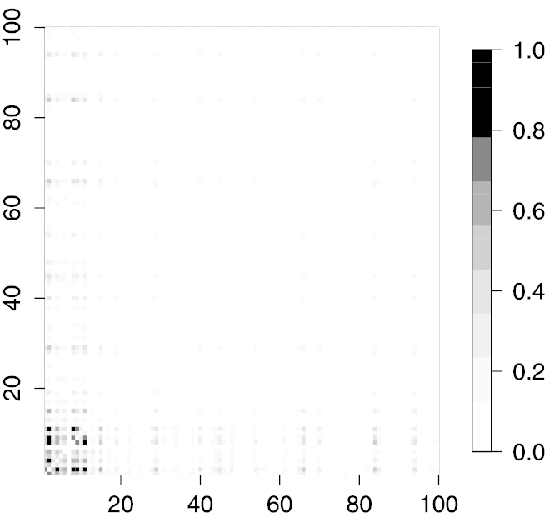}
 & \includegraphics{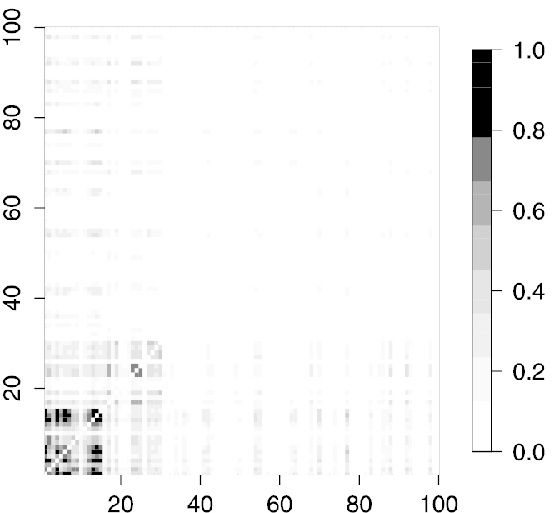}\\
\footnotesize{(a)} & \footnotesize{(b)}
\end{tabular}
\caption{Heatmaps of the matrix $\bar{\bfA}$ using data sets with \textup{(a)}
$p=150$, 0\% experiment dependency, one functional group, subsampling
level 70\% and $(\lambda_1,\lambda_2)=(9,9)$; \textup{(b)} $p=300$, 0\%
experiment dependency, two functional groups, subsampling level 70\%
and $(\lambda_1,\lambda_2)=(9,15)$. For clarity, only the first
$100\times100$ entries are shown and the functional groups are placed
at positions 1--15 and 16--30, respectively.}
\label{figmatA}
\end{figure}

\subsubsection{Estimated \texorpdfstring{$\bar{\mathbf{A}}$}{barA} and tuning parameter selection}
Figure~\ref{figmatA} shows the heatmaps of the matrix $\bar{\bfA}$ for
two data sets with different numbers of functional groups. For visual
clarity, the genes are ordered according to their true group
memberships. In both cases, the matrix demonstrates a clear block
structure. In particular, in the two-group case both pathways are
visible, although the first one is more prominent. We remark here that
the difference in signal strength between the two pathways is
introduced by chance variation during data generation and the use of
subsampling is necessary for the identification of the weaker group.
Although we present results obtained with a subsampling level of 70\%,
a range of reasonable subsampling levels can be chosen without
significantly affecting the final results (supplementary information
Section~7 [\citet{supp}]). The other tuning parameter $\bolds\lambda$ is chosen such
that the matrix $\bar{\bfA}$ displays optimal contrast between the
pathway and nonpathway groups, and we shall use this as guidance for
assessing the quality of $\bar{\bfA}$ and selecting $\bolds\lambda$.

\begin{figure}
\centering
\begin{tabular}{@{}cc@{}}

\includegraphics{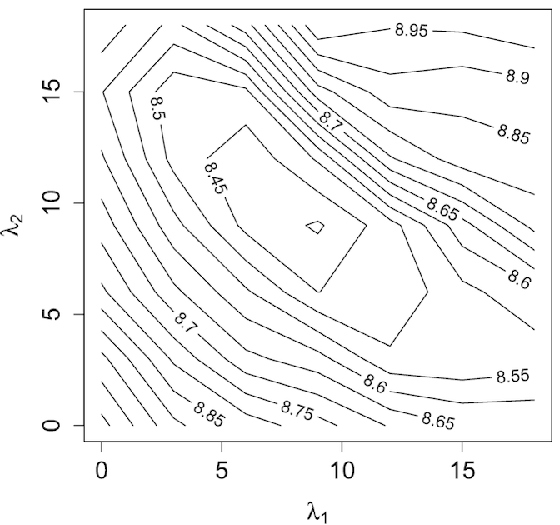}
 & \includegraphics{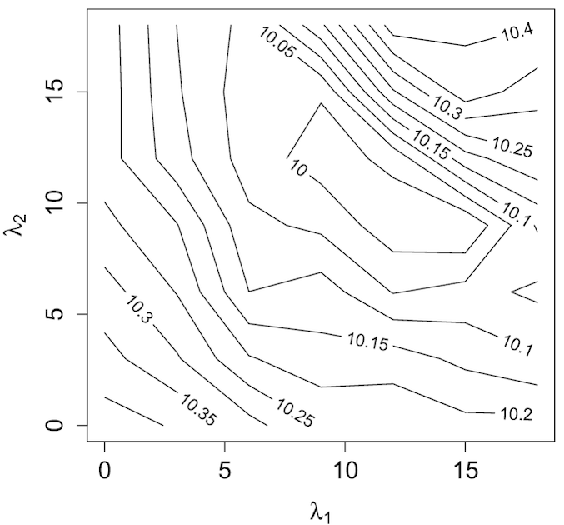}\\
\footnotesize{(a)} & \footnotesize{(b)}
\end{tabular}
\caption{Contour plots of the entropy of the upper triangular entries
of $\bar{\bfA}$ on the grid $(\lambda_1,\lambda_2)\in\{0,3,\ldots
,18\}^2$
using data sets with \textup{(a)} $p=150$, 0\% experiment dependency, one
functional group and subsampling level 70\%; \textup{(b)} $p=300$, 0\% experiment
dependency, two functional groups and subsampling level 70\%.}
\label{fig_contour}
\end{figure}

Among the common approaches for the selection of optimal tuning
parameters, cross-validation-based methods are used in \citet
{Waaijenborg:2008}, \citet{Parkhomenko:2009} and \citet{Lee:2011}.
However, all of their methods involve dividing a sample into multiple
sets, which is impractical for data sets with only a few tens of
observations. \citet{Witten:2009b} proposed an alternative
permutation-based method which estimates the $p$-value of the maximal
correlation found by performing SCCA on permuted samples. Due to the
large number of partitions and subsamplings required in our method,
this approach would be very computationally expensive. Instead we
measure the effectiveness of $\bolds\lambda$ using the entropy of
$\bar
{\bfA}$, defined as
%
\begin{equation}
H(\bfA)=-\sum_{i<j,\bfA_{ij}>0} (\bfA_{ij}/S_{\bfA})
\log(\bfA _{ij}/S_{\bfA}), \label{eq_entropy}
\end{equation}
where $S_{\bfA}=\sum_{i<j}\bfA_{ij}$. The entropy quantifies the
sharpness of its distribution and thus is indicative of the signal
intensity. Figure~\ref{fig_contour} plots the contours of $H(\bar
{\bfA})$
for the same two data sets used in Figure~\ref{figmatA}. Regions with
low entropy correspond to~$\bolds{\lambda}$, leading to a matrix with
better signal intensity.

%
%

%
%

\begin{figure}
\centering
\begin{tabular}{@{}cc@{}}

\includegraphics{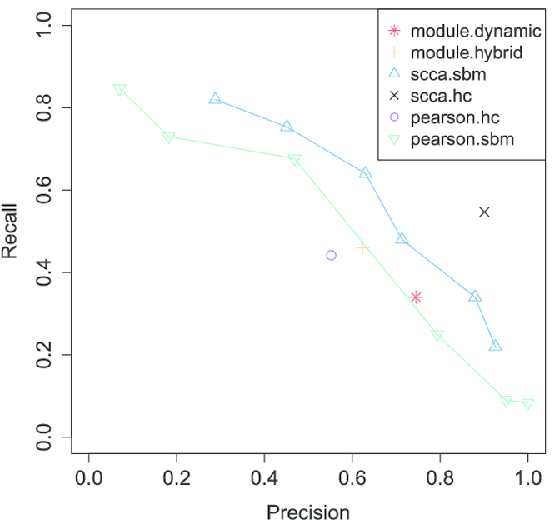}
 & \includegraphics{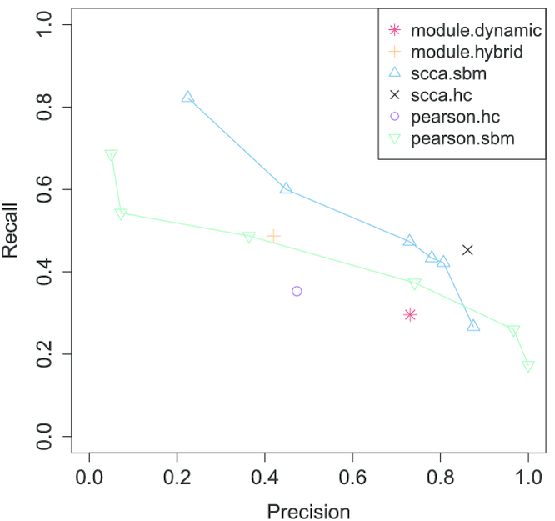}\\
\footnotesize{(a)} & \footnotesize{(b)}
\end{tabular}\vspace*{3pt}
\begin{tabular}{@{}c@{}}

\includegraphics{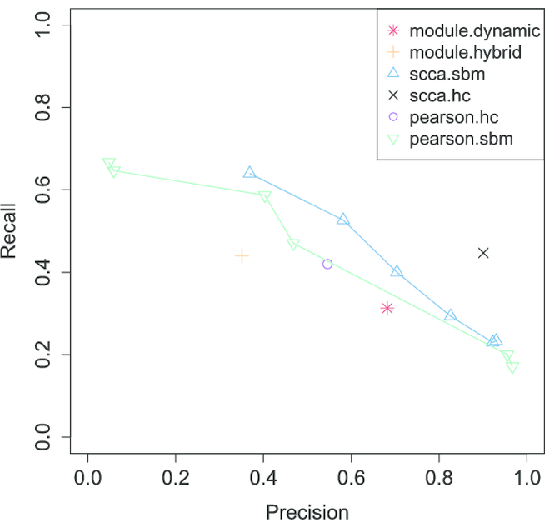}
\\
\footnotesize{(c)}
\end{tabular}
\caption{Classification performance of different methods using data
sets with $p=500$, one pathway group, subsampling level 70\%, and \textup{(a)}
0\%, \textup{(b)} 33\% and \textup{(c)} 67\% of experiment dependency. \textit
{pearson.sbm} and \textit{scca.sbm} are applied to matrices at
discretization levels $\{0.3,0.4,\ldots,0.8\}$ (from left to right on
the curve).}
\label{fig_comp1c}
\end{figure}

\subsubsection{Performance comparison}
Figure~\ref{fig_comp1c} compares the classification performance of our
methods, \textit{scca.sbm} and \textit{scca.hc}, with four
correlation-based methods, \textit{pearson.hc}, \textit{pearson.sbm},
\textit{module.dynamic} and \textit{module.hybrid}. The methods are
named by cross-mixing the following to allow for comparisons in the
two-stage procedure:

\textit{scca}: Calculate $\bar{\bfA}$'s with $\bolds{\lambda}\in
\{
9,12,\ldots,27\}^2$ and select 10 of these with the smallest entropy
values. The final cluster membership (after community detection) is
decided by a majority vote based on the selected $\bar{\bfA}$'s, so
only stable clusters and cluster members are chosen.

\textit{pearson}: Pearson's correlation matrix after the data is
normalized using equation~\eqref{eq_normalized_by_Knorm} and \textit
{Knorm} estimates.

\textit{module}: Transformed Pearson's correlation matrix used in
\citet
{Langfelder:2007}.

\textit{sbm}: Fit a SBM on a discretized edge weight matrix (at level
$\{0.3,0.4,\ldots,\break 0.8\}$) using the unconditional pseudo-likelihood
algorithm in \citet{Amini:2013} with $Q=2$ (or 3) initialized by
spectral clustering with perturbation. Select the cluster with the
highest internal connectivity based on the estimates.

\textit{hc}: HC with Ward's distance and cut the dendrogram when
clusters of size less than 25 start to appear as the number of clusters
$Q$ increases. The choice of this upper bound is based on the size of
the cluster selected in \textit{scca.sbm}, and a range of reasonable
numbers can be used without affecting the final results.

\textit{dynamic}, \textit{hybrid}: HC with dendrogram cutting methods
in the R package \texttt{dynamicTreeCut} [\citet{Langfelder:2008}].

%

Figure~\ref{fig_comp1c} plots the average \textit{precision} and
\textit
{recall} of the above six methods calculated on 10 simulation data sets
for each level of experiment dependency. It can be seen that using our
SCCA approach to compute edge weights in general leads to higher
\textit
{precision} across all experiment dependency levels. Of the two ways of
community identification, \textit{scca.hc} produces higher \textit
{precision} than \textit{scca.sbm} at comparable \textit{recall} levels.

\begin{table}
\caption{Classification performance of different methods using data
sets with $p=500$, two pathway groups, subsampling level 70\%, and
various levels (0\%, 33\% and 67\%) of experiment dependency}
\label{table_2c}
\begin{tabular*}{\textwidth}{@{\extracolsep{\fill}}lcccccc@{}}
\hline
& \multicolumn{2}{c}{\textbf{0\%}} & \multicolumn{2}{c}{\textbf{33\%}} &
\multicolumn
{2}{c@{}}{\textbf{67\%}}\\[-6pt]
& \multicolumn{2}{c}{\hrulefill} & \multicolumn{2}{c}{\hrulefill} &
\multicolumn
{2}{c@{}}{\hrulefill}\\
& \textbf{Precision} & \textbf{Recall} & \textbf{Precision} &
\textbf{Recall} & \textbf{Precision} &
\textbf{Recall}\\
\hline
& \multicolumn{6}{c@{}}{Pathway 1}\\
\textit{scca.hc} & 0.861 & 0.533 & 0.831 & 0.441 & 0.811 &
0.433\\
\textit{pearson.hc} & 0.238 & 0.233 & 0.497 & 0.427 & 0.471 &
0.393\\
\textit{module.dynamic} & 0.718 & 0.3\phantom{00} & 0.742 & 0.333 & 0.764 & 0.38\phantom{0}\\
\textit{module.hybrid} & 0.439 & 0.407 & 0.544 & 0.447 & 0.453 &
0.385\\[3pt]
& \multicolumn{6}{c}{Pathway 2}\\
\textit{scca.hc} & 0.808 & 0.487 & 0.890 & 0.489 & 0.833 &
0.420\\
\textit{pearson.hc} & 0.438 & 0.387 & 0.323 & 0.307 & 0.460 &
0.273\\
\textit{module.dynamic} & 0.758 & 0.4\phantom{00} & 0.808 & 0.347 & 0.8\phantom{00} &
0.4\phantom{00}\\
\textit{module.hybrid} & 0.565 & 0.473 & 0.529 & 0.387 & 0.455 &
0.46\phantom{0}\\
\hline
\end{tabular*}
\end{table}

Table~\ref{table_2c} shows the same performance measures obtained from
data sets containing two independent functional groups for \textit
{scca.hc}, \textit{pearson.hc}, \textit{module.dynamic} and \textit
{module.hybrid}. The numbers are averages from 10 simulation data sets
for each level of experiment dependency. Similar to the one-group case,
we choose the smallest $Q$ that produces two clusters of size less than
25 as the cutoff in HC. We remark here that when multiple groups are
present, \textit{scca.sbm} tends to detect only the strongest signal
group while failing to pick up the weaker one. This can be explained by
considering the within-class homogeneity assumption in the SBM model
and noting that the degree distribution is often less homogeneous in
the weaker signal group (see, e.g., Figure~\ref{figmatA}). The
conditional pseudo-likelihood algorithm in \citet{Amini:2013} is also
not sensitive enough to detect the finer distinctions. Results from
\textit{pearson.sbm} are also omitted as they are very noisy. In all
the cases, \textit{scca.hc} demonstrates the best \textit{precision} at
comparable, if not better, \textit{recall}.


\subsection{Application to real data}
\label{subsec_realData}
We tested the performance of our procedure by applying it to \textit
{Arabidopsis thaliana} microarray expression data retrieved from
AtGenExpress 
(\url{http://www.arabidopsis.org/servlets/TairObject?type=expression\_set\&id=1007966941}). The analyzed data set included expression measurements
collected from shoot tissues subject to oxidation stress for 22,810
genes under 13 experiment conditions with two replicates for each
experiment. In these experiments, the plants were treated with methyl
viologen (MV), which led to the formation of reactive oxygen species
(ROS). Various studies have shown that depending on the type of ROS, a
different biological response is provoked. Thus, by focusing on the ROS
induced by MV, we were able to show and validate that the results of
our pathway gene search were supported, in part, by other already
published ROS-related microarray experiments.

A subset of all 22,810 genes was selected for analysis based on the
following criteria. (i) The experiment variance of the gene exceeds
0.1. An unvarying expression profile suggests the gene has an activity
level unaltered by the particular stress condition, and hence is
unlikely to be part of any stress-induced pathway. The inclusion of
such genes may cause problems in covariance estimation as well. We also
removed genes with a suspiciously high experiment variance, as it could
suggest inaccuracy in measurements. (ii) The discrepancy between the
two replicates is smaller than 2 for each experiment. This ensures only
genes with consistent measurements are included in our analysis. (iii)
The minimum expression level exceeds 7. More active genes are likely to
possess stronger signals, making our search easier. This requirement
further trims down the data set to a smaller size more desirable for
our procedure. We note here that the inclusion of (iii) is optional---if
running time is not a concern, the minimum expression level could be
either lowered or entirely removed. The final subset for analysis
contained 2718 genes.

Potential functional groups were found by \textit{scca.hc}. Due to the
complexity and noise level of the data set, we did not expect the
entropy~\eqref{eq_entropy} to have a clean-cut unimodal distribution.
Furthermore, the presence of many groups with varying signal strengths
implies each may need a different optimal $\bolds\lambda$ for
detection. For example, strong groups are likely to require more
regularization or, in other words, larger $\bolds\lambda$. For
this reason, we performed our search in multiple stages starting from
large $\bolds\lambda$ for stronger groups to smaller
$\bolds
\lambda$ for weaker ones. At every stage, the groups found were removed
from the original set before proceeding to the next stage. The upper
bound on $\bolds\lambda$ was found by increasing $\bolds
\lambda$ until the entropy stabilized. Searching down from this upper
bound, we chose $\bolds\lambda$ from three grids: $\{90,100,110\}
^2$, $\{60,70,80\}^2$ and $\{30,40,50\}^2$. The cutoff level $Q$ in HC
was increased incrementally until at least five clusters of size less
than 30 appeared. A reasonable range of numbers can be used to choose
the cutoff and our results are not very sensitive to the choice of this
number. The full procedure produced 13 groups of genes, the full list
of which, including annotations, can be found in Section~6 of the
supplementary information [\citet{supp}].

To test the biological significance of all 13 groups found (i.e.,
whether there is a functional relationship between genes within the
various groups), we first examined for enrichment of gene product
properties, collectively designated gene ontology (GO) annotations,
within each group using information available at The Arabidopsis
Information Resource 
(\url{http://www.arabidopsis.org/tools/bulk/index.jsp}). We determined that
8 out of 13 groups were highly enriched with genes having the same GO
annotation and calculated their $p$-values using Fisher's exact test to
compare with the counts obtained from the full analyzed data set
(Table~\ref{tab_GO}).

%
\begin{table}
\tabcolsep=0pt
\centering
\caption{GO enrichment of groups}
\label{tab_GO}
\begin{tabular*}{\textwidth}{@{\extracolsep{\fill}}lccc@{}}
\hline
 &  & \textbf{Number of genes with}  & \\
\textbf{Group ID}&\textbf{Enriched GO term}&\textbf{enriched terms}&
$\bolds{P}$\textbf{-values}\\
\hline
\phantom{0}1 & Chloroplast organellar gene & 10 out of
15{\scriptsize{\tabnoteref{tbl1}}} & $1.10\times
10^{-4}$\\
\phantom{0}2 & Phenylpropanoid-flavonoid biosynthesis & 3 out of 4 & $6.65\times
10^{-7}$\\
\phantom{0}3 & Glucosinolate biosynthsis & 7 out of 7 & $1.95\times10^{-14}$\\
\phantom{0}4 & Chloroplast organellar gene & 3 out of 3 & $7.83\times10^{-3}$\\
\phantom{0}5 & Ribosome & 10 out of 15 & $7.20\times10^{-13}$\\
\phantom{0}8 & Ribosome & 5 out of 6 & $8.31\times10^{-8}$\\
10 & Photosystem I or II & 8 out of 10 & $2.87\times10^{-14}$\\
12 & Endomembrane system & 3 out of 4 & $2.35\times10^{-3}$\\
\hline
\end{tabular*}
\tabnotetext[1]{tbl1}{4 out of the 10
chloroplast genes are mitochondrial organellar genes.}
\end{table}

In addition to the GO enrichment approach for validating the groups,
and in order to support the biological significance of the groups
found, we also evaluated other forms of evidence. We were able to
determine that for several groups the genes placed in the groups encode
for known pathways. For example, group 2 genes encode steps in the
phenylpropanoid-flavonoid (FB) biosynthesis pathway, and group 3 genes
encode for steps in the glucosinolate (GSL) biosynthesis pathway. Both
are well-studied secondary metabolic pathways. Flavonoids are compounds
of diverse biological activities such as anti-oxidants, functioning in
UV protection, in defense, in auxin transport inhibition and in flower
coloring [\citet{Gachon:2005,Naoumkina:2010,Taylor:2005,Woo:2005}],
and GSLs are sulfur-rich amino acid-containing compounds which become
active in response to tissue damage and are believed to offer a
protective function [\citet{Sonderby:2010,Verkerk:2009,Yan:2007}]. A
considerable number of genes in both pathways are induced by broad
environmental stresses and regulated at the transcriptional level.
Based on the lists of genes associated with these two pathways reported
in \citet{Kim:2012}, our analyzed data set contained 13 FB pathway genes
and 26 GSL pathway genes. The precisions of our search are 75\% and
100\%, respectively.

In order to assess the likelihood that genes in the remaining groups
could also encode steps within specific pathways, we reviewed
microarray data from plants subjected to other forms of oxidative
stress (these experiments are similar to the experiment from which our
data set using MV was obtained). Using this approach we found that
genes in each of the additional seven groups (1, 4, 5, 8, 9, 11, 12)
were strongly associated in these independent experiments
(supplementary information Section~6 [\citet{supp}]).

Of all the groups found, groups 6, 7 and 13 remain uncharacterized in
the literature. Nonetheless, using CoExSearch [part of the ATTD-II
database (\url{http://atted.jp/top\_search.shtml} \#CoexVersion)], all four
genes in group 7 were correlated to some degree with abiotic stress
conditions. We also found these genes were common anoxia-repressed
genes [\citet{Loreti:2005}]. The lack of complete characterization for
these groups in the current literature leaves potential scope for
further biological examination.

For comparison we applied \textit{pearson.hc}, \textit{module.dynamic}
and \textit{module.hybrid} to the same data. As the simulation study
suggests the latter two methods in general have better performance than
\textit{pearson.hc}, particularly in the multi-group case, we will
present the results from these two methods and refer to Section~6 in
the supplementary information [\citet{supp}] for \textit{pearson.hc}-based results. In
order to compare with our results, we chose two cuts of the dendrogram
such that the first cut produced the same number of groups as our
method, and the second one led to groups with sizes comparable to ours.
The first cut resulted in 13 groups with sizes ranging from 60 to 293.
We picked the three most promising groups based on their annotations
and the GO analysis is summarized in Table~\ref{tab_GO_shallow}.
Although all of them have statistically significant $p$-values, their
precisions are quite low. In particular, group 11 contains our group 2
as a subset and includes 11 genes (out of 76) in the FB pathway and 5
genes are in the isoprenoid biosynthesis pathway. These two pathways
are derived from different initial precursors and are known to be
unrelated. We note here that at this cut level, the GSL pathway cannot
be identified by the method. The second cut produces 66 groups with
sizes from 5 to 81. We picked five small groups for analysis and only
one group with genes localized in chloroplast has significant GO
enrichment (Table~\ref{tab_GO_deep}). Even so, these genes are unlikely
to be functionally related. The comparison suggests our method can
achieve better precision and lead to more biologically meaningful
groupings of genes.

\subsection{Effects of tuning parameters}
\label{sec_sensitivity}
To systematically study the effects of different tuning parameters on
the identification of gene functional groups, we perform sensitivity
analysis for different choices of subsampling levels and penalty
parameter $\bolds{\lambda}$ using both the simulated and real data
discussed above. For the sake of completeness, we also compare tuning
parameters from the HC and SBM procedures. Overall, our results are
reasonably stable for a range of $\bolds{\lambda}$ values. Further
stability can be achieved by pooling results from different $\bolds
{\lambda}$. As expected, the choice of subsampling level is more
important when there exist multiple functional groups. Our results
suggest levels between 50\% and 80\% can all be considered in practice.
For community detection, HC is more robust than SBM in the sense that
the classification results are not sensitive to the cutoff chosen. The
results are summarized in Section~7 of the supplementary information [\citet{supp}].

\begin{table}
\tabcolsep=0pt
\caption{GO enrichment of groups---first cut}
\label{tab_GO_shallow}
\begin{tabular*}{\textwidth}{@{\extracolsep{\fill}}lccc@{}}
\hline
 &  & \textbf{Number of genes with}  & \\
\textbf{Group ID}& {\textbf{Enriched GO term}}&\textbf{enriched terms}&$\bolds{P}$\textbf{-values}\\
\hline
\phantom{0}9 & Cell wall & 16 out of 81 & $4.46\times10^{-6}$\\
10 & Defense response & 29 out of 78 & $1.58\times10^{-2}$\\
11 & Phenylpropanoid-flavonoid biosynthesis & 11 out of 76 &
$5.42\times10^{-12}$\\
\hline
\end{tabular*}
\end{table}

\begin{table}[b]
\caption{GO enrichment of groups---second cut}
\label{tab_GO_deep}
\begin{tabular*}{\textwidth}{@{\extracolsep{\fill}}lcc@{}}
\hline
 &  & \textbf{Number of genes with}  \\
\textbf{Group ID}& {\textbf{Enriched GO term}}&\textbf{enriched terms}\\
\hline
62 & NA & 0 out of 6 \\
63 & Chloroplast & 4 out of 6 \\
64 & Located in plasma membrane & 2 out of 5 \\
65 & Located in plasma membrane & 3 out of 5 \\
66 & Pyridoxine biosynthetic process & 2 out of 5 \\
\hline
\end{tabular*}
\end{table}

\section{Discussion}
\label{sec_discussion}
In this paper we focus on the problem of estimating gene group
interactions in gene networks, where data are given in the form of
nodes and their associated covariates and estimation of the true
network is a challenging task. We propose\vadjust{\goodbreak} a new method to construct an
edge weight matrix for the full network by applying SCCA to sampled
subsets of genes with random partitioning. To evaluate the quality of
the constructed network, subsequent analysis of the community
structures is applied to identify potential gene functional groups.
Although the work is presented under the setting of gene networks, we
believe our approach can be generally applicable to answer similar
questions in other biochemical networks and even networks in other
fields that are sparse and have similar covariate features.

Compared to other popular ways of measuring gene interactions, our SCCA
approach is more conceptually appealing. By seeking maximally
correlated sets of genes among randomly sampled subsets, this approach
provides an aggregated measure of gene partial correlations when the
correct conditional set is unknown, and thus gives us a better chance
of capturing group interactions. As demonstrated in both simulation and
real data applications, one of the main attractions of our procedure is
its high \textit{precision}. Although it does not seem to greatly
improve \textit{recall}, this is not a huge drawback in light of the
search algorithm by \citet{Kim:2012}. Given the accuracy of our search
results in general, one can use these identified genes as ``seed genes''
to initiate a more complete search and expand on the current lists.

Our approach can be modified to handle other practical situations. When
it is known in advance that some genes operate in the same functional
group, one may incorporate the prior knowledge by lowering the
penalties associated with those genes in the SCCA algorithm. Although
we have focused on the case with disjoint functional groups, our method
of constructing an edge weight matrix is still applicable to the
overlapping case as long as the shared genes possess strong direct or
partial interactions with all the other functional genes (supplementary
information Section~5 [\citet{supp}]). However, a different community detection method
[e.g., mixed membership SBM; \citet{Airoldi:2008}] should be applied to
identify the overlapping structures.

The core of our procedure consists of an implementation of SCCA by
LASSO regression, and this naturally opens room for further
investigation. For example, it would be interesting to find out if
using other penalty functions yields different results, more
importantly, whether SCCA can be implemented using a different
optimization criterion or a more efficient algorithm to lessen the
computational cost of our procedure. 
In the theoretical aspect, it would be desirable to incorporate
sparsity into our asymptotic analysis.

On the community detection side, although we used SBM and HC as
examples, there are many other available methods to be further
explored, especially their properties in relation to the edge weight
matrix $\bar{\bfA}$. The use of SBM and HC also gives rise to other
interesting extensions. As noted in Section~\ref{subsec_simulation},
conventional SBM does not perform well when there are multiple groups,
which is mainly caused by the heterogeneity of node degrees.\vadjust{\goodbreak} However,
fitting a degree-corrected model using the conditional
pseudo-likelihood algorithm does not seem to offer significant
improvement. It would be desirable to carry out further study on the
theoretical properties of the degree-corrected SBM and characterize its
identifiability problem. Another possible extension is to modify these
algorithms to take weighted adjacency matrices without discretization.
Developing a practical but more systematic way of choosing the cutoff
level for HC also invites future study.


\begin{supplement}[id=suppA]
\stitle{Supplementary information}
\slink[doi,text={10.1214/14-AOAS792SUPP}]{10.1214/14-AOAS792SUPP} 
\sdatatype{.pdf}
\sfilename{aoas792\_supp.pdf}
\sdescription{As\-ymptotic analysis and additional explanations of the
procedure, additional simulation and real data results. The code for
estimating the edge weight matrix can be requested from \printead*{e5}.}
\end{supplement}

%

%

\printaddresses
\end{document}